\documentclass[a4paper, 10pt]{article}

\usepackage{multicol}
\usepackage{datetime}
\usepackage{sectsty}
\usepackage{graphicx}
\usepackage{fancyhdr}
\usepackage{xifthen}
\usepackage{wrapfig}
\usepackage{caption}
\usepackage{setspace}
\usepackage{xcolor}
\usepackage{pifont}

\sectionfont{\fontsize{12}{12}\selectfont}
\subsectionfont{\fontsize{11}{11}\selectfont}
\newcommand{\twentyeighteen}[1]{} 
\newcommand{\twentynineteen}[1]{}

\newdateformat{monthyeardate}{\monthname[\THEMONTH], \THEYEAR}
\begin{document}
\onehalfspacing

\centerline{\large{\textbf{Urban Socio-Technical Systems: An Autonomy and Mobility Perspective}}} 
\centerline{} 
\centerline{Weizi Li} 
\centerline{University of Memphis} 
\centerline{wli@memphis.edu} 
\centerline{} 
\par

\subsection*{Introduction} 
The future of the human race is urban. The world's population is projected to grow an additional 2.5 billion by 2050, with \emph{all} expected to live in urban areas~\cite{un}. This will increase the percentage of urban population from 55\% today to 70\% within three decades and further strengthen the role of cities as the hub for information, transportation, and overall socio-economic development. 
Unlike any other time in human history, the increasing levels of autonomy and machine intelligence are transforming cities to be no longer just human agglomerations but a fusion of humans, machines, and algorithms making collective decisions, thus \textbf{complex socio-technical systems}. 
My passion is to study, design, and develop cities through a systems lens and make them \textbf{intelligent, equitable, resilient}, and \textbf{sustainable}. My current research focuses on \textbf{urban autonomy and mobility}---an essential infrastructure of urban socio-technical systems and on which all social sectors depend. 
Specifically, my ongoing efforts can be summarized in three categories: \ding{172} shared autonomy between humans and robotic systems, \ding{173} robust intelligent transportation systems and city-scale traffic, and \ding{174} urban informatics and network science.

\vspace{-.5em}
\subsection*{Shared Autonomy between Humans and Robotic Systems}   
\begin{figure}[ht]
	\centering
	\includegraphics[width=\linewidth]{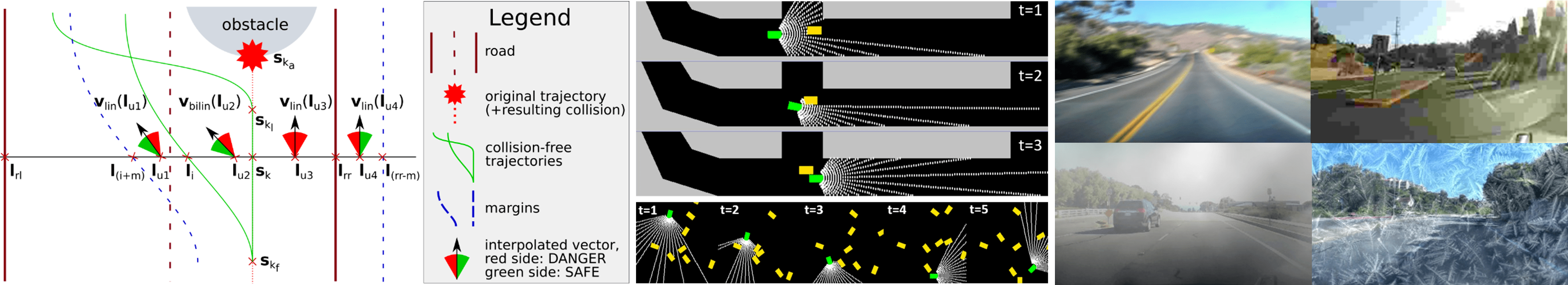}
	\vspace{-2em}
	\caption{\small{Autonomous vehicle (AV) control and navigation. Left: learning safe maneuver via trajectories from physics-based simulations. Middle: swift maneuver of AV (green) in mixed traffic via improved inverse reinforcement learning and trust-region optimization. Right: sample images for testing robust autonomous steering.}} 
	\vspace{-1.25em}
	\label{fig:av}
\end{figure}

\vspace{-.5em}
One of the most exciting innovations in urban mobility is autonomous driving.
While it carries the potential to revolutionize our transport systems, the convergence of technology, infrastructure, and policy for full adoption of autonomous vehicles (AVs) may still require decades to come.
Mixed traffic, where humans share space and interact with robot vehicles at various autonomy levels, will be the dominant transport mode in the near future. 
To better prepare for mixed traffic, we need to increase the capabilities of robot vehicles.  
One promising learning approach for autonomous driving is imitation learning (IL) via human demonstrations.
However, a drawback of IL is covariate shift, which can cause the learnt policy to function poorly in novel environments~\cite{dagger}. 
To address this issue, we develop IRL-HC~\cite{iros}---a platform that employs inverse reinforcement learning (IRL) to train AVs in mixed traffic. 
IRL-HC consists of both 3D simulation and 2D simulation: the 3D simulation generates driving scenarios and collects data, while the 2D simulation serves as the `expert' to analyze and resolve any collision by planning alternative trajectories for the AV (see Fig.~\ref{fig:av} Left).
IRL-HC addresses a fundamental limitation of the original IRL~\cite{irl}, that is, imposing a uniform prior on all features can cause inferior imitation performance of the learner to the expert even if their feature expectations resemble each other. 
IRL-HC adopts a non-uniform prior on task features 
so that both expert demonstrations and domain knowledge can be incorporated into learning.  
As a result, the AV trained using our method drives the longest distance collision-free compared to other state-of-the-art (SOTA) techniques under the same vehicle speed. 
Some driving scenarios are shown in Fig.~\ref{fig:av} Middle. 
\textbf{IRL-HC~\cite{iros} and its precedent ADAPS~\cite{icra} are among the few frameworks that adopt physics-based simulation for training safe autonomous driving}.

\begin{figure}[ht]
	\centering
	\includegraphics[width=\linewidth]{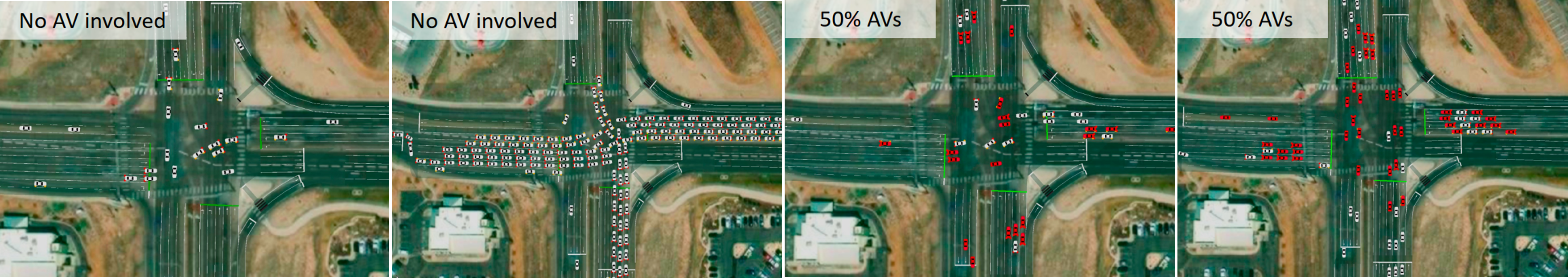}
	\vspace{-2em}
	\caption{\small{Mixed traffic control and coordination. AVs are denoted in red. Left to Right: traffic at signalized intersection; traffic at unsignalized intersection, congestion develops within 15 mins; mixed traffic at signalized intersection; mixed traffic at unsignalized intersection, no congestion is observed after 15 mins.}} 
	\vspace{-2em} 
	\label{fig:mix}
\end{figure}

Recent studies~\cite{flow} have shown that a fleet of AVs can be adopted to control mixed traffic through their interactions with nearby human-driven vehicles. 
The modeling and control of mixed traffic is challenging due to the absence of effective models and the fact that traditional control theory is rooted in model-based design. 
This motivates the use of reinforcement learning---a model-free technique that can be applied to high-dimensional, multi-agent control tasks. 
We choose intersections to study the potential of mixed traffic control since they are the flow hubs of a road network and carry unique challenges including varied topology and conflicting traffic streams.
We propose an ego-global reward function that considers the waiting
time and queue length of each connecting road at the intersection, and a coordination mechanism to resolve the conflicting traffic streams within the intersection~\cite{intersection}.
Our results show that using 40\% or more AVs in traffic, we can achieve the same throughput compared to traffic lights (see Fig.~\ref{fig:mix}); and using 100\% AVs, we can reduce the waiting time of all vehicles at the intersection up to 75\%. 
\textbf{Our work is the first to study the control and coordination of mixed traffic at real-world, complex intersections}.


\vspace{-.5em}

\subsection*{Robust ITS and City-Scale Traffic} 
\vspace{-.5em}
In the past decade, intelligent transportation systems (ITS) applications at all scales (from individual vehicles to city-scale traffic) have gained significant improvements thanks to the advancement of machine learning. 
While effectiveness and efficiency are apparent measures of an application, robustness is also crucial: an application that is vulnerable to even small perturbations has limited use in practice.
In the following, we introduce two of our studies that explore model robustness, and our effort on one of the fundamental tasks of ITS---traffic reconstruction. 

So far, AV training and testing are mainly conducted in ideal environments without any adversarial conditions.  
Daily driving, however, will experience all types of environment conditions (e.g., lighting and weather). 
The external factors in conjunction with internal factors of sensors can lead to quality-varying input data and subsequently affect the control performance.   
We propose a gradient-free adversarial training technique, named AutoJoin~\cite{autojoin}, for robust imaged-based steering. 
The core idea is to use a decoder attachment to the
regression model creating a denoising autoencoder within the architecture so that the tasks `autonomous steering' and `denoising sensor input' can be
jointly learnt and reinforce each other’s performance. 
Compared to other SOTA methods with testing on over 5M perturbed (samples are shown in Fig.~\ref{fig:av} Right) and clean images, AutoJoin
achieves up to 40\% performance gain under gradient-free perturbations while improving on clean performance up to 300\%. 
AutoJoin is also very efficient as it saves up to 83\% time per training epoch and 90\% training data over other SOTA techniques. 
To the best of our knowledge, \textbf{AutoJoin is the most effective and efficient gradient-free adversarial training technique for imaged-based autonomous steering}.

\begin{figure}[ht]
	\centering
	\includegraphics[width=\linewidth]{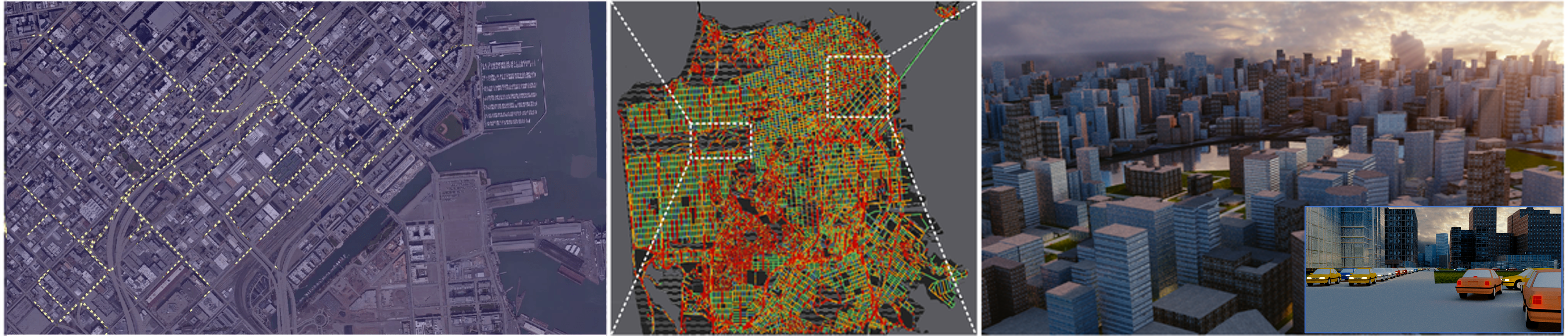}
	\vspace{-2em}
	\caption{\small{City-wide traffic reconstruction and visualization of San Francisco. Left to Right: reconstruction results shown in 2D vehicle motions, congestion states, and 3D virtual downtown, respectively. }}
	\vspace{-.75em}
	\label{fig:city}
\end{figure}

A major goal of ITS is to alleviate traffic congestion.
To understand traffic jams, we need to reconstruct traffic at the city scale to analyze congestion causes and identify network bottlenecks. 
Mobile data such as GPS traces are the most promising source for the task because of their broader coverage of a city. 
However, mobile data are usually incomplete and sparse, thus requiring several pre-processing steps before they can be used for reconstruction: 1) map-matching, which maps noisy mobile data onto road segments and infers the traversed path of a vehicle; 2) travel-time estimation, which estimates the travel time of road segments; and 3) missing-data estimation, which interpolates spatial and temporal missing measurements. 
We improve map-matching and travel-time estimation achieving up to 97\% improvement in estimation accuracy compared to existing techniques~\cite{iet,itsm}. 
For missing-data estimation, we develop a simulation-based technique that employs metamodel for the task while ensuring traffic flow consistency at the boundaries of areas with and without data~\cite{siga}. 
The reconstruction results are showcased in Fig.~\ref{fig:city}. 
Our approach is \textbf{the first and remains the state-of-the-art to reconstruct citywide traffic dynamics via incomplete and sparse mobile data}.

\vspace{-.5em}

\begin{figure}[ht]
	\centering
	\includegraphics[width=\linewidth]{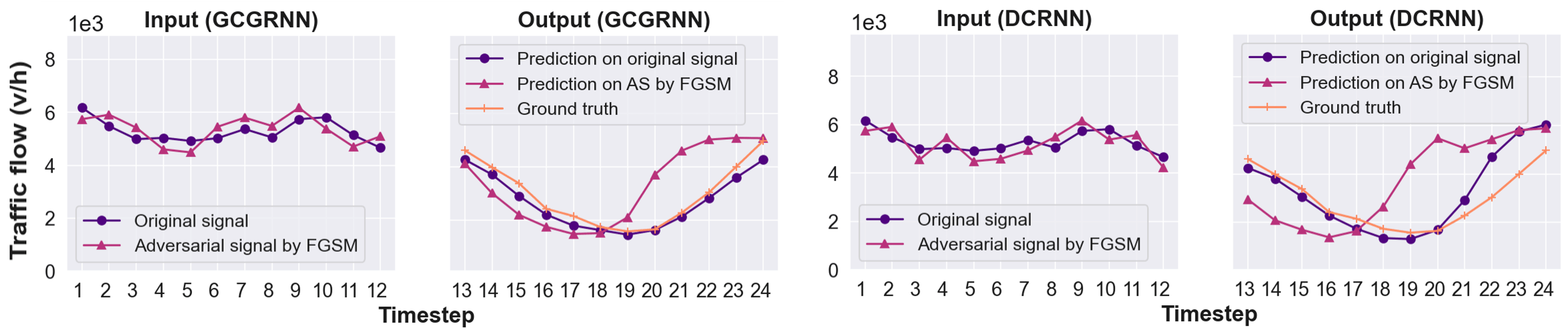} 
	\vspace{-2em}
	\caption{\small{Adversarial attacks on traffic state prediction. GCGRNN and DCRNN are latest prediction models, and FGSM is an attack algorithm. The adversarial signals have significant (negative) impact on prediction accuracy. }} 
	\vspace{-1.25em}
	\label{fig:attack}
\end{figure}

Following reconstruction, another important task for traffic control and management is traffic state prediction. 
Most existing studies focus on improving prediction accuracy without exploring model robustness. 
We propose an adversarial attack framework to test the robustness of traffic state prediction models~\cite{blackbox}. 
The target model (the model being tested) is treated as a black-box. 
No knowledge of the model architecture, training data, and parameters is assumed, but the adversary can feed the target model with arbitrary input and obtain output. 
The input-output pairs can then be used in training a substitute model to mimic the target model's behavior and, subsequently, to generate adversarial signals.  
We test our adversarial attack framework on two SOTA traffic state prediction models and find that we can degrade the model performance up to 54\% using the adversarial signals. 
The results are demonstrated in Fig.~\ref{fig:attack}.
\textbf{Our work is the first to conduct black-box adversarial attack on traffic state prediction models}.

\vspace{-.5em}
\subsection*{Urban Informatics and Network Science} 
\begin{figure}[ht]
	\centering
	\includegraphics[width=\linewidth]{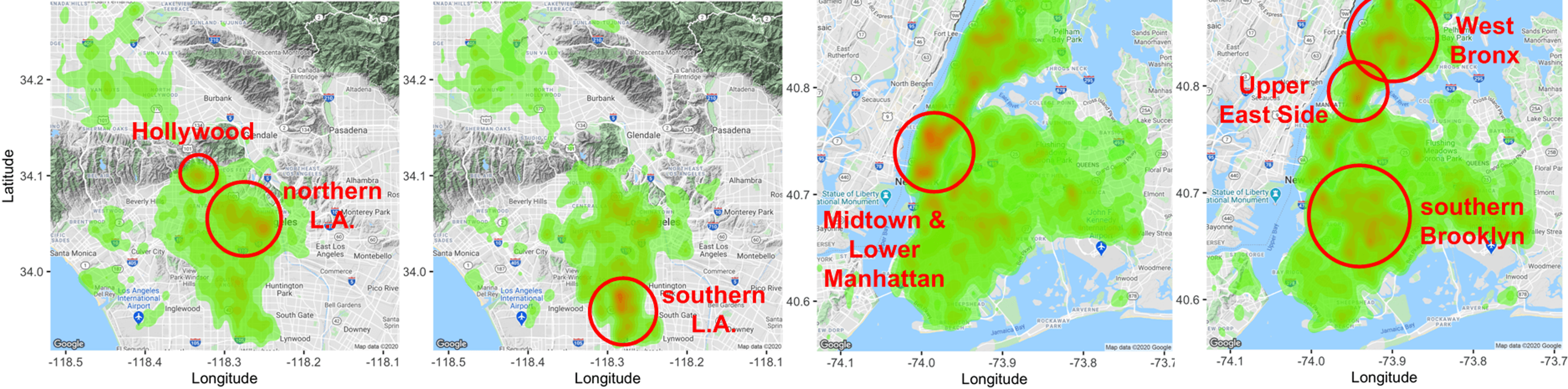}
	\vspace{-2em}
	\caption{\small{Comparing the hot spots of traffic accidents before and after the pandemic lockdown policy (Mar, 2020). In LA, the hot spots shift from Hollywood and northern LA to southern LA; in NYC, the hot spots shift from Midtown \& Lower Manhattan to other locations. The shift destinations tend to be lower-income areas.}}
	\vspace{-.75em}
	\label{fig:accident}
\end{figure}

\vspace{-.5em}
Besides traffic congestion, traffic safety is the most critical topic in transportation-related urban informatics. 
Since 2020, the pandemic has caused a dramatic change to human mobility. 
We investigate the impact of the pandemic and subsequent mobility changes on traffic safety~\cite{accident}.
Using accident data from Los Angeles and New York City, we find that the impact is not merely a blunt
reduction in traffic and accidents; rather, 1) the proportion of accidents unexpectedly increases for Hispanic and Male
groups; 2) shown in Fig.~\ref{fig:accident}, the hot spots of accidents have shifted from higher-income areas
(e.g., Hollywood and Lower Manhattan) to lower-income areas (e.g., southern LA and southern Brooklyn); 3) the severity level
of accidents decreases with the number of accidents regardless of transportation modes. Understanding those variations of
traffic accidents not only sheds a light on the unbalanced impact of the pandemic across demographics and geographic factors,
but also helps policymakers and planners design effective and equitable policies and interventions during emergent events.  
\textbf{Our work is the first using mathematical models and statistical tests to analyze the impact of the pandemic on traffic safety}. 

The explosive growth of Internet, networking technology, and ubiquitous sensors have resulted in the creation of numerous graphs with increasing complexity. 
Urban environments are no exception; from informatics to physical devices, we are connected more than ever. 
The interconnectedness and interdependence of emerging graphs can cause enormous resources for processing, storage, communication, and decision-making on these graphs. 
This emphasizes the necessity of graph sparsification: an edge-reduced graph of a similar structure
to the original graph is produced while various user-defined
graph metrics are largely preserved. Existing graph sparsification
methods are mostly sampling-based, which introduces high
computation complexity in general and lack of flexibility for different reduction objectives. We present SparRL~\cite{sparrl}, \textbf{the first
generic and effective graph sparsification framework enabled
by deep reinforcement learning}. SparRL can easily adapt to
different reduction goals and promises graph-size-linear
complexity. Extensive experiments show that SparRL outperforms
all prevailing sparsification methods in producing high-quality
sparsified graphs concerning a variety of objectives~\cite{sparrl}.

In addition to the example work mentioned above, we have conducted a plethora of studies on vehicle learning and control~\cite{dc,attention}, ITS~\cite{gcgrnn,survey,compressed,metro,bike}, and multi-agent systems and virtual humans~\cite{individual,insects,memory,commonsense,apprentice,distribution,purpose}.

\subsection*{Future Work} 
\vspace{-.5em} 
In the future, I plan to explore the above research categories in depth, as well as other urban-related topics. 

\textbf{Shared Autonomy between Humans and Robotic Systems}.  
Mixed traffic control is still at early stages. There are many exciting directions to pursue based on our intersection work~\cite{intersection}.    

\vspace{-\topsep}
\begin{itemize}
\setlength{\parskip}{0pt} \setlength{\itemsep}{0pt plus 1pt}
    \item How robust is a policy to its input observations? 
    What are the minimal observations needed for an effective policy? 
    How about a different observation format, e.g., policy from bird's-eye view images?
    \item How can we generalize an effective policy at one intersection to other intersections? 
    How can we coordinate adjacent intersections? 
    What are the necessary communication mechanisms needed? 
    \item What novel learning algorithms can we develop to facilitate shared autonomy? What is the best way to incorporate human data into learning under a shared autonomy setting?
    \item What insights can we draw for shared autonomy of other sectors such as manufacturing, agriculture, and healthcare? Can we develop a general policy learning framework across different domains?  
\end{itemize}
\vspace{-\topsep}

\textbf{City-scale Traffic and Robust ITS}.    
Modeling citywide traffic is challenging but necessary for relieving traffic congestion. 
I am interested in extending our reconstruction study~\cite{siga} in two aspects.  

\vspace{-\topsep}
\begin{itemize}
\setlength{\parskip}{0pt} \setlength{\itemsep}{0pt plus 1pt}
    \item Efficiency improvement. Many advanced ITS applications such as digital twins require highly efficient reconstruction. 
    I plan to use hybrid simulation (macroscopic + microscopic) instead of just microscopic simulation
    for improving the reconstruction efficiency. 
    \item Efficiency-fidelity trade-off. Different ITS applications have different requirements on efficiency and reconstruction fidelity. 
    A systematic and quantitative report that explores efficiency-fidelity trade-off in various ITS applications can benefit many research communities.   
    I plan to fill this void. 
\end{itemize}
\vspace{-\topsep}

Robust training will always be a crucial topic in developing intelligent systems. 
I will continue exploring it for ITS applications at all levels. 
At the microscopic level, I will keep improving model robustness in both perception and control for autonomous driving;
at the mesoscopic level, I will examine policy robustness at various road-network structures such as roundabouts and other junctions; 
at the macroscopic level, I plan to develop both accurate and robust city-scale traffic prediction and optimization techniques.     

\textbf{Urban Informatics and Network Science}. 
I will continue investigating traffic-related urban informatics with a focus on the \emph{equity} aspect of transportation systems.
One idea is to extend our traffic safety work~\cite{accident} by incorporating the study of infrastructure: roads and bridges are known to connect affluent sectors while excluding the poor~\cite{equal}. 
Regarding network science, an immediate plan is to extend our SparRL framework~\cite{sparrl} to cover more types of graphs, including directed and weighted graphs, so that it can support more applications. 
Other plans include improving its scalability to accommodate very large graphs and extending it to other network-related tasks such as link prediction and label classification. 

My ultimate goal is to imbue cities with intelligence, equity, resilience, and sustainability.  
I am eager to explore other urban topics under these four themes, in addition to the topics mentioned above.  
In the future,
regarding \textbf{intelligence}, I would like to apply machine learning (especially reinforcement learning) to urban planning and policy development;
regarding \textbf{equity}, I plan to explore equitable access of transportation and the allocation of resources; 
regarding \textbf{resilience}, I want to design infrastructure (transportation and beyond) that is robust under extreme weather and temperature;
with additional research on energy and food \textbf{sustainability}, collectively, I hope to develop urban to be a warm home for future humanity. 
	
\begin{small}

\end{small}

\end{document}